\documentclass[aps,twocolumn,showpacs,citeautoscript,reprint,pre,floatfix]{revtex4-1}

\usepackage{bm}
\usepackage{graphicx}
\usepackage{amsmath}
\usepackage{amsfonts}
\usepackage{amssymb}
\usepackage{siunitx}
\usepackage{color}
\usepackage{hyperref}

\definecolor{Blue}{rgb}{0.3,0.3,0.9}
\definecolor{Red}{rgb}{0.9,0.3,0.3}
\definecolor{Green}{rgb}{0.3,0.6,0.3}
\definecolor{Black}{rgb}{0.0,0.0,0.0}

\newcommand{\revision}[1]{\textcolor{Black}{{#1}}}

\begin{document}

\title{Spin-dependent electronic lenses based on hybrid graphene nanostructures}

\author{Yuriko Baba}

\email{yuribaba@ucm.es}

\affiliation{GISC, Departamento de F\'{\i}sica de Materiales, Universidad Complutense, E--28040 Madrid, Spain}

\author{Marta Saiz-Bret\'{\i}n}

\affiliation{GISC, Departamento de F\'{\i}sica de Materiales, Universidad Complutense, E--28040 Madrid, Spain}


\begin{abstract}
 
We study electronic transport in graphene/ferromagnetic insulator hybrid devices. The system comprises an armchair graphene nanoribbon with a lens-shaped EuO ferromagnetic insulator layer deposited on top of it. When the device supports a large number of propagating modes, the proximity exchange interaction of electrons with the magnetic ions of the ferromagnetic insulator results in electrons being spatially localised at different spots depending on their spin. We found the spin-dependent electron focusing is robust under moderate edge disorder. A spin-polarised electric current can be generated by placing a third contact in the proper place. This opens the possibility to use these effects for fabricating tunable sources of polarized electrons.

\end{abstract}

\maketitle

\section{Introduction}   \label{sec:intro}

Graphene is a truly two-dimensional material with remarkable electronic properties. Soon after the discovery of graphene, Morozov \emph{et al.}\ found extremely low electron-phonon scattering rates that set the fundamental limit on possible charge carrier mobilities at room temperature~\cite{Morozov08}. Moreover, Bolotin \emph{et al.}\ observed ballistic transport in ultraclean suspended samples up to $\SI{2}{\micro \meter}$ at cryogenic temperatures, suggesting long coherence lengths.~\cite{Bolotin08} Coherent transport has already been studied and observed in graphene-based nanodevices. In this context, Mu\~{n}oz-Rojas \emph{et al.}\ found numerically that coherent transport through graphene nanoconstrictions is also robust with respect to variations of constriction geometries and edge defects~\cite{Munoz-Rojas06}. This theoretical finding has been recently proved in experiments in graphene nanoconstrictions at low temperature~\cite{Caridad18,Clerico18}.

\begin{figure}[htb]
\centerline{\includegraphics[width=1\columnwidth]{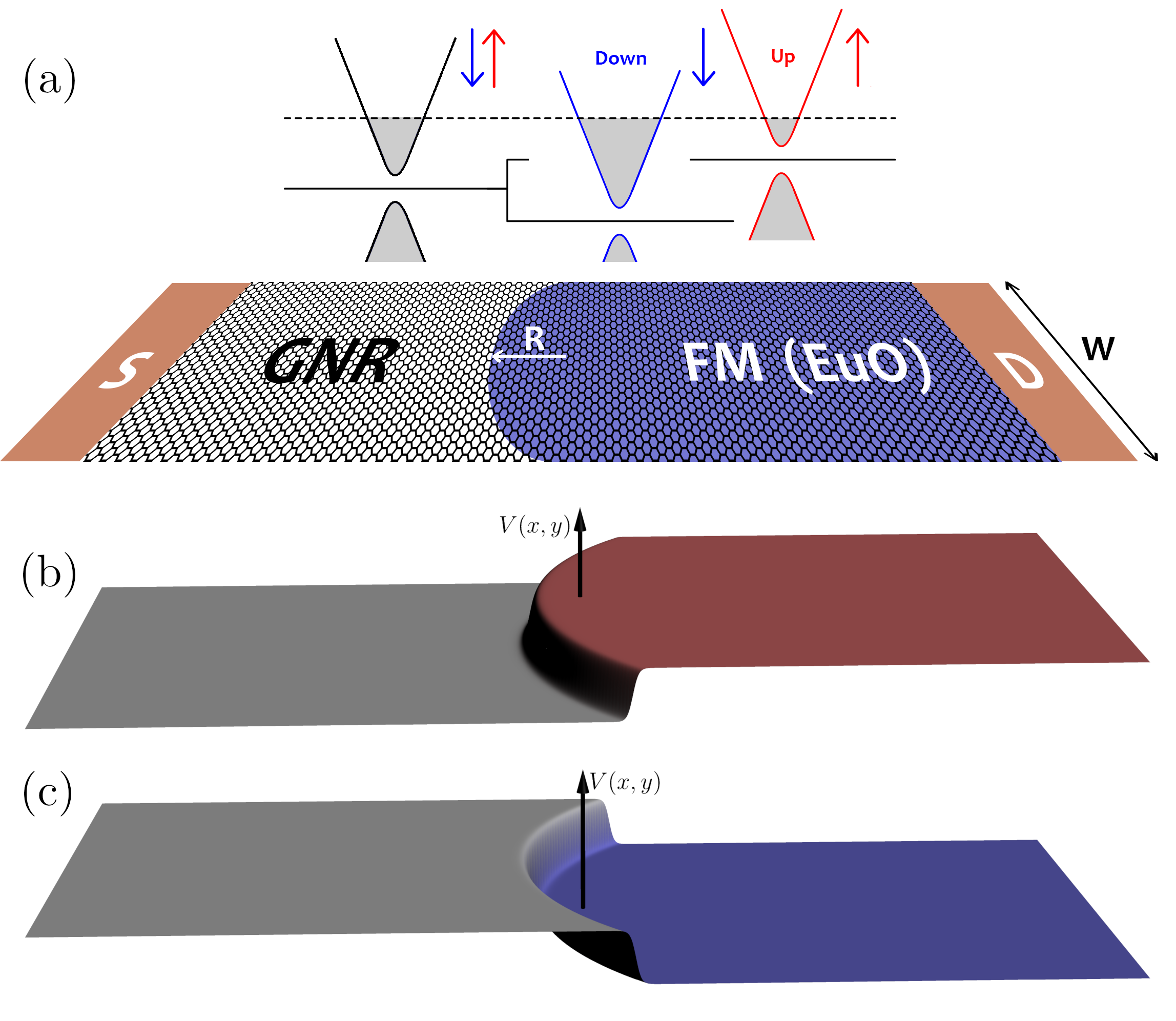}}
\caption{(a)~Schematic view of the device. The semiconducting armchair graphene nanoribbon is connected to source~(S) and drain~(D) leads, with a ferromagnetic insulator lens grown on top (shown as the blue region in the figure). The proximity exchange interaction between magnetic ions in the EuO layer and charge carriers in the GNR induces a Zeeman-like splitting. (b) Spin-up electrons interact with a lens-shaped potential barrier while (c)~spin-down electrons are exposed to a lens-shaped potential well.}
\label{fig1}
\end{figure}

These properties pave the way to exploit interference effects of coherent electron transport, making graphene-based devices ideal candidates to become the building blocks for future carbon-based electronic circuits. In particular, massless Dirac electrons of graphene in ballistic regime behave in many ways similar to photons~\cite{Bhandari18}. Electronic junctions (\textit{p-n} junctions) take advantage of the optical-like electron dynamics to achieve electron guiding and focusing over distances exceeding $\SI{1}{\micro\meter}$~\cite{Rickhaus2013}. An applied magnetic field can help to probe the optical-like nature of electrons in \textit{p-n} junctions and to design positive and negative refraction lenses~\cite{Chen16}. Recently, Bai \emph{et al.} have realized nanoscale \textit{p-n} junctions with atomically sharp boundaries in graphene, enabling the observation of quantum interference patterns~\cite{Bai18}. Electron collimation in a Hall bar with added zig-zag contacts has been imaging with a liquid-He cooled scanning gate microscope by Bhandari \emph{et al.}, demonstrating that ray-tracing simulation agrees with the experimental images~\cite{Bhandari18}. The achievement of the aforementioned junctions led to the experimental realization of a wide range of theoretically proposed lenses, which include flat and curved lenses in sheets and nanoribbons. A flat \textit{p-n} junction generates the focalization of a point source in an only one focus in both graphene sheets \cite{Cheianov2007} and nanoribbons \cite{Libisch2017}. Furthermore, electron beams can be controlled by parabolic lenses \cite{Liu2017}, which collimate the beam, or by circular \textit{p-n} junctions that lead to the focus of electrons when arranged as a big scattering region in a sheet~\cite{Cserti2007} or in arrays in a nanoribbon~\cite{Tang2016}.

The coherent behavior of electrons in ballistic graphene suggests interesting applications in spintronics as well. \revision{In Ref.~\cite{Moghaddam2010} the case of a planar lens based on a ferromagnetic stripe deposited on bulk graphene is proposed, showing that an unpolarized electronic beam can be collimated with a finite spin polarization.}
By the other hand, hybrid nanostructures containing EuO ferromagnetic insulator layers deposited on top of graphene quantum rings~\cite{Munarriz12,Saiz-Bretin15} and superlattices~\cite{Munarriz13,Diaz14} provide a route to design spin-filters and spin-valves. The ferromagnetic layers induce a proximity exchange splitting of the electronic states in graphene~\cite{Haugen08,Swartz12}, resulting in the appearance of a spin-dependent potential profile that it is analogous to an induced Zeeman splitting of the energy levels. Similar effects are predicted to occur in silicene-based devices~\cite{Nunez16,Nunez17}. 

\revision{In contrast to the graphene sheets studied in Ref.~\cite{Moghaddam2010},}
in this work we study a new design of a hybrid nanostructure based on a graphene nanoribbon (GNR) with a lens-shaped EuO ferromagnetic insulator layer deposited on top of it. The GNR is attached to two non-magnetic contacts in a standard two-terminal configuration. 
\revision{The de Broglie wavelength corresponding to the energies considered in this work is larger than the typical size of the system proposed. Even if we are away from the optical regime, interesting features can be still obtained. In fact, we will demonstrate that the quantum nature of electrons in the device enables spin-dependent electron focusing.}
The interference pattern of electrons depends on the number of modes entering the system as well as the electron energy. In the one-mode region the interference pattern is uniform and can be explained solely by the transmission coefficient. On the contrary, the interference pattern is richer in wide GNRs, when a larger number of modes enter the system. In this case we observe a spatial separation of the electrons according to their spins. Then one could have a spin-polarised electric current by placing a third contact in the proper place. Finally, we demonstrate the spin-dependent electron focusing is robust under moderate edge disorder.

\section{Model}   \label{sec:model}

The hybrid system under study consists of an armchair GNR of width $W$, connected to source and drain leads, on top of which there is a EuO ferromagnetic insulator lens of radius $R$, as shown schematically in figure~\ref{fig1}(a). 
Source and drain leads are taken as semi-infinite GNRs. To model the system, we consider a tight-binding Hamiltonian of a single electron in the $\pi$-orbitals of graphene within the nearest-neighbor approximation
\begin{equation}
\mathcal{H}_{\sigma}=-t\sum_{\langle i,j\rangle}|i\rangle\langle j|
+ \sigma\,\Delta_\mathrm{ex}\sum_{i\in {\mathcal L}}|i \rangle\langle i|\ .
\end{equation}
The site energy is set to zero without losing generality. Here $|i \rangle$ is the ket vector of the atomic orbital of the $i$th carbon atom and $t=\SI{2.7}{\electronvolt}$ is the hopping energy between neighboring atoms $i$ and $j$. Ab initio calculations yield values of the order of 
$\SI[input-quotient=:, output-quotient=\text{ -- }]{100:300}{\milli\electronvolt}$ 
for the proximity exchange interaction energy $\Delta_\mathrm{ex}$ for graphene in close proximity to chalcogenides (EuO and EuS)~\cite{Hallal17}. Throughout this work we fix
$\Delta_\mathrm{ex}=\SI{200}{\milli\electronvolt}$, with $\sigma=\pm 1$ for spin-up and spin-down electrons. The exchange interaction is induced only at the atoms that are in direct contact with the lens-shaped ferromagnetic layer (the full set of them is labeled as $\cal L$ in the above equation). We assume that electrons are in the fully coherent regimen and travel ballistically across the system. The quantum transmission boundary method~\cite{Lent90,Ting92}, combined with the effective transfer matrix method~\cite{Schelter10}, is used to compute wave functions and the spin-dependent transmission coefficient $\tau_{\sigma}(E)$ as a function of energy (see Ref.~\cite{Munarriz14} for further details). Because the proximity exchange interaction has the characteristic length scale of one atomic layer, the splitting is induced only in the regions of the GNR which are just below the ferromagnetic lens. 
\revision{Therefore, no spacial smoothness in the potential profile needs to be considered.} For the chosen system geometry, a spin-up (spin-down) electron propagating along the sample will be subject to a two-dimensional lens-shaped potential barrier (well), as shown in panels~(b) and~(c) of figure~\ref{fig1} respectively.

\section{Results}   \label{sec:results}

\begin{figure}[t!]
\centerline{\includegraphics[width=0.85\columnwidth]{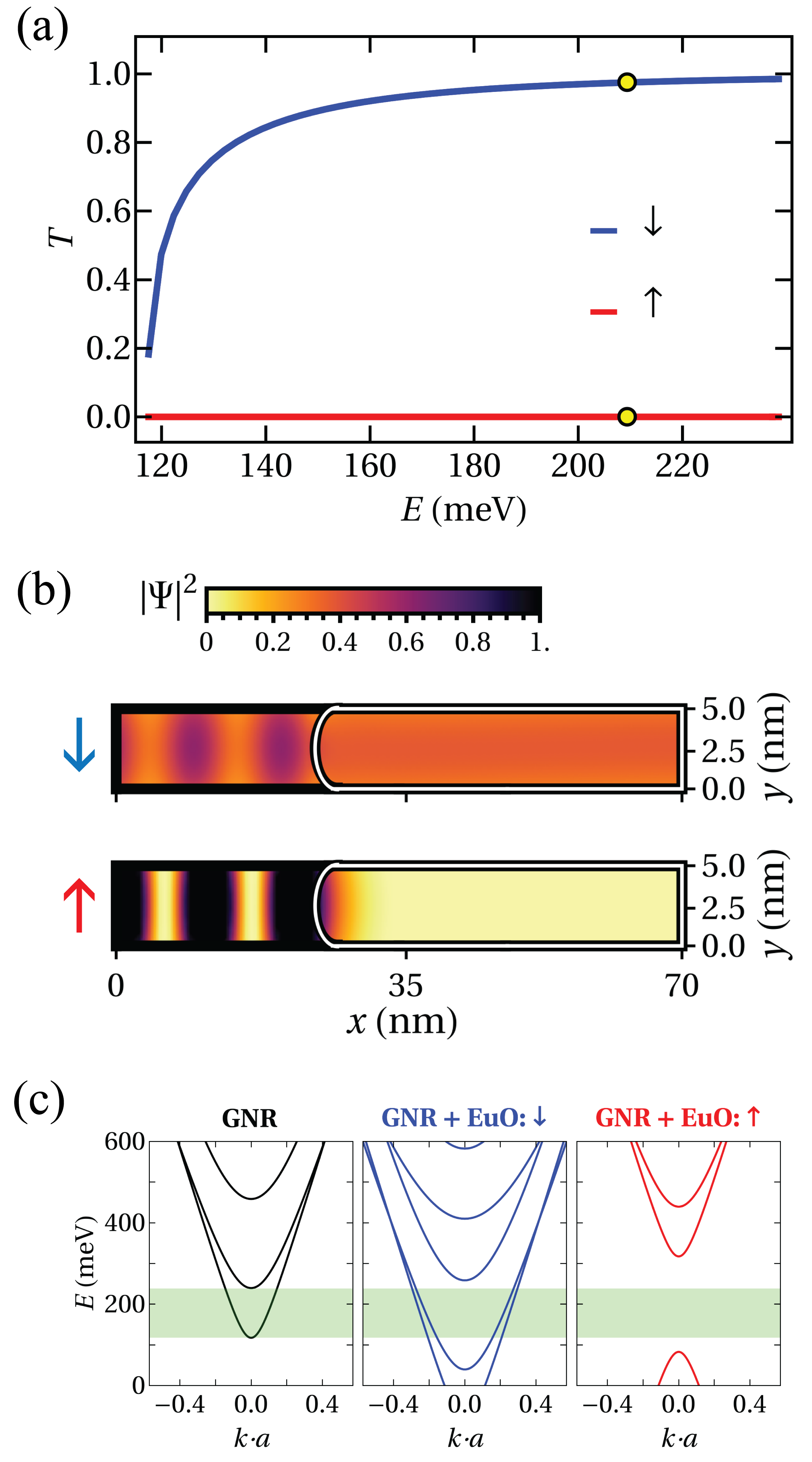}}
\caption{(a)~Transmission coefficient for both spins in the one mode energy regime when $W=\SI{5}{\nano\meter}$, (b)~square modulus of the envelope wave function for both spins at the energy marked with yellow circles above and (c)~band structure (the shadowed area indicates the one mode energy range for a GNR in the absence of ferromagnetic lens).}
\label{fig2}
\end{figure}

The armchair GNRs considered in this work are semiconducting and the energy gap is controlled by their width $W$. First we focus on the interplay between the exchange energy $\Delta_\mathrm{ex}$ and the GNR width. To do so, we fix the radius of the ferromagnetic lens, $R=W/2$, and analyze the transmission coefficient in the one-mode energy regime, as well as the wave functions and the band structure. When the width of the GNR lies approximately within the range $\SI{2.5}{\nano\meter} < W < \SI{6.5}{\nano\meter}$, the transmission coefficient increases with energy for one of the spins (spin down), while for the other one (spin up) remains vanishingly small, as shown in figure~\ref{fig2}(a). In figure~\ref{fig2}(b) we plot the square modulus of the envelope wave function for the energy marked with yellow circles in figure~\ref{fig2}(a). We can observe that for spin up electrons, the wave function is reflected at the ferromagnetic lens, leading to zero transmission. This can be understood if we look at the band structure for this GNR shown in figure~\ref{fig2}(c). When the ferromagnetic lens is placed on top of the GNR, the whole band structure shifts towards higher (lower) energy for spin up (down). For GNRs with $\SI{2.5}{\nano\meter} < W < \SI{6.5}{\nano\meter}$, the energy range considered (corresponding to the one-mode regime for a GNR without exchange interaction) falls in the energy gap for one of the spins, giving rise to a highly polarized electron transmission. The same behavior is observed regardless of the the specific shape of the ferromagnetic layer (with a curvature or planar) and we conclude that narrow GNRs can behave as very efficient spin filters.

\begin{figure}[t!]
\centerline{\includegraphics[width=0.85\columnwidth]{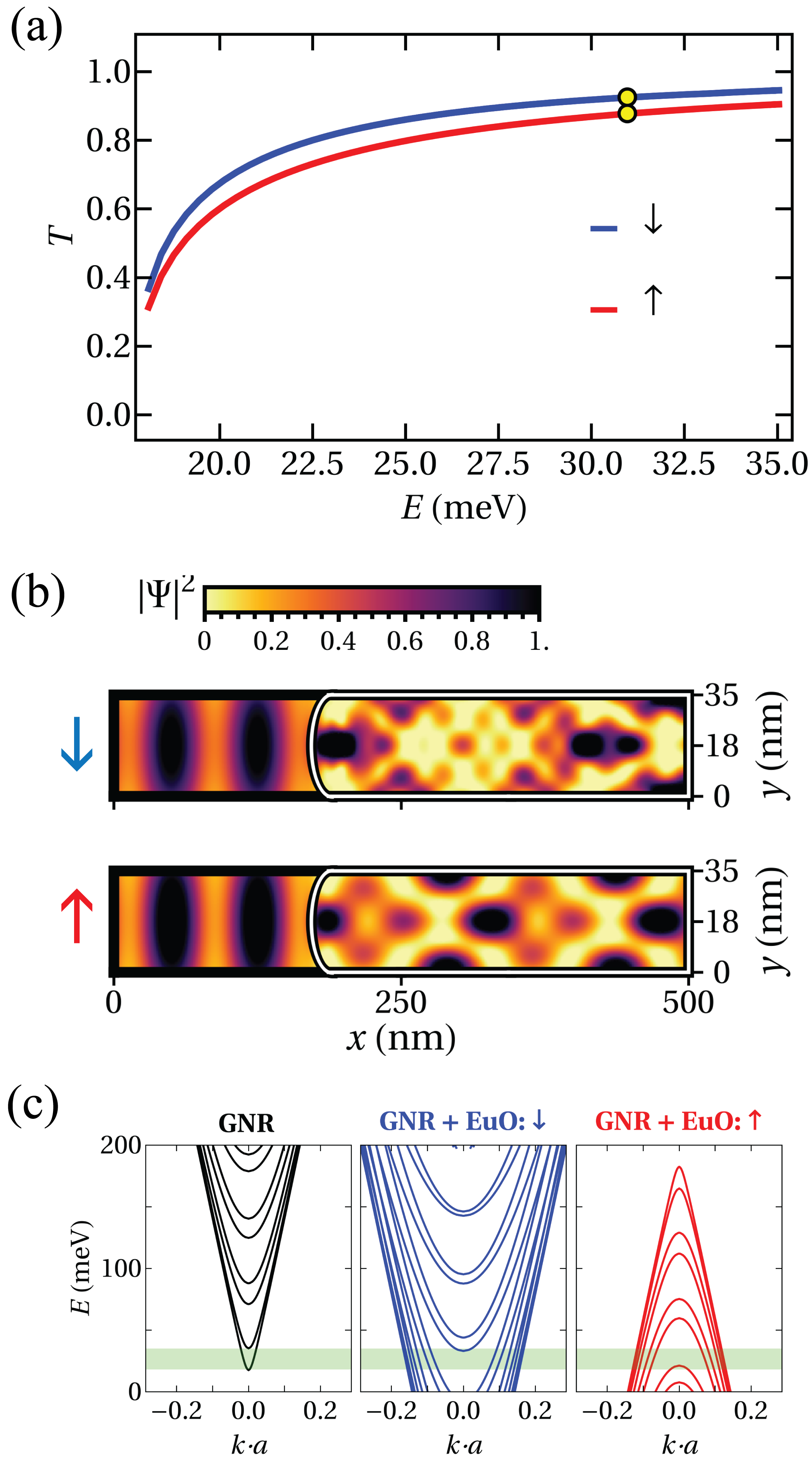}}
\caption{(a)~Transmission coefficient for both spins in the one mode energy regime when $W=\SI{35}{\nano\meter}$, (b)~square modulus of the envelope wave function for both spins at the energy marked with yellow circles above and (c)~band structure (the shadowed area indicates the one mode energy range for a GNR in the absence of ferromagnetic lens).}
\label{fig3}
\end{figure}

However, as the GNR becomes wider both spins contribute to the electron transmission and more energy modes come into play when the ferromagnetic lens is placed [see figure~\ref{fig3}(c)]. The superposition of several modes for both of the spins along with the lens curvature lead to multiple spots where a certain spin is focused, especially at the edges and the center of the GNR as can be observed in figure~\ref{fig3}(b) for a GNR of width $W=\SI{35}{\nano\meter}$. Ultimately, these focus regions could lead to a spin-polarised electric current by placing a third non-magnetic contact in the proper place, close to the edge of the ferromagnetic layer.

\subsection{Spin Polarization}

\begin{figure}[htb]
\centerline{\includegraphics[width=0.8\columnwidth]{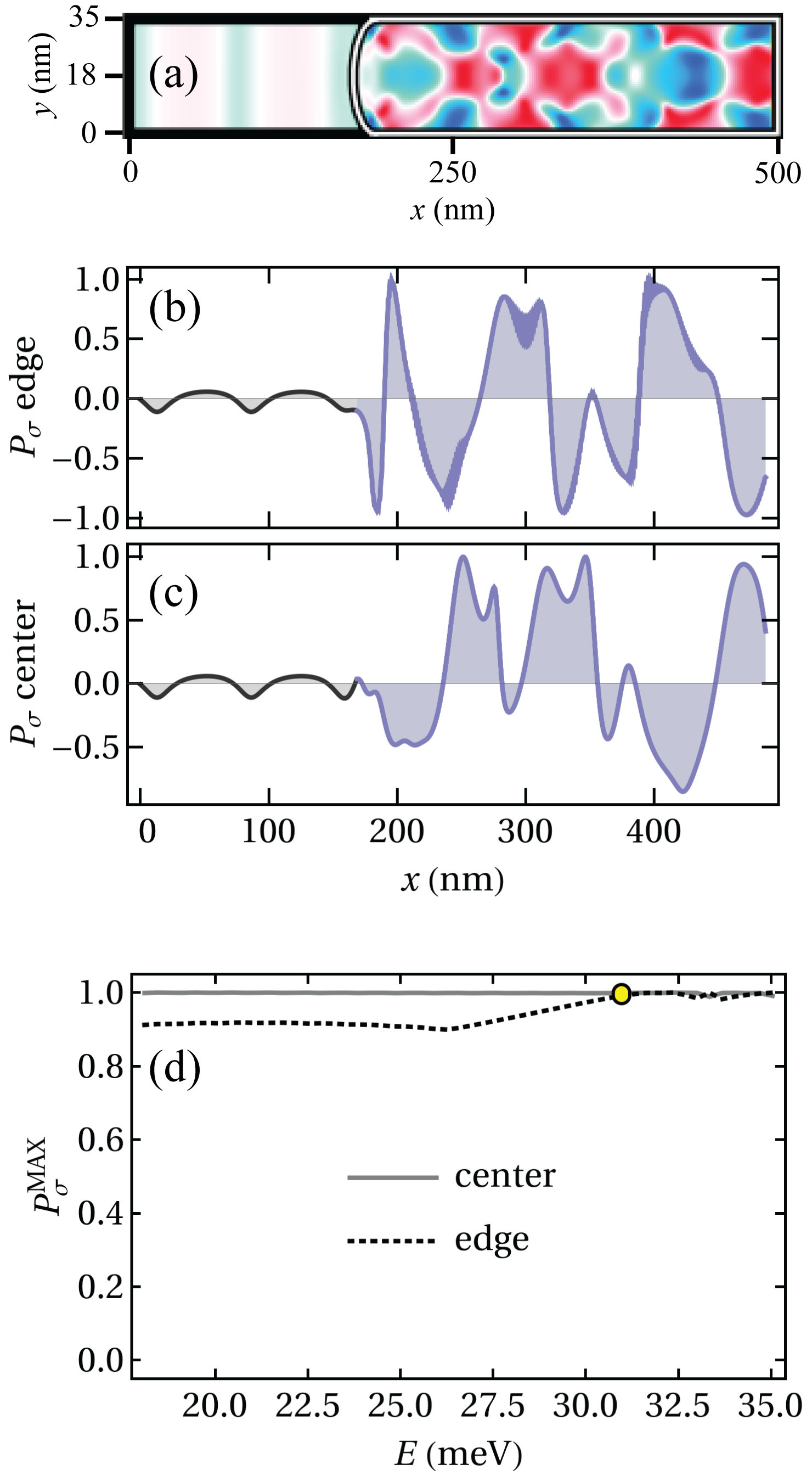}}
\caption{(a) Density plot of the spin polarisation for a GNR of width $W=\SI{35}{\nano\meter}$ and  radius $R=W/2$ (the same GNR considered in figure~\ref{fig3}). Blue (red) area represents a high magnitude of the spin down (up) density. (b)~Spin polarisation along the center of the GNR (c)~and the edge. The blue color represents the polarisation in the lens region. (d) Maximum of the spin polarisation at the center and at the edge for an energy window within the one mode regime. The yellow circle indicates the energy for which the spin polarisation has been plotted in the upper panels.}
\label{fig4}
\end{figure}

To better visualize the focus spots for each spin, we define the spin polarisation as 
\begin{equation}
P(E)=\frac{|\psi_{\uparrow}(E)|^2-|\psi_{\downarrow}(E)|^2}{|\psi_{\uparrow}(E)|^2+|\psi_{\downarrow}(E)|^2}\ .
\label{polarisation}
\end{equation}
It will be used as figure of merit to assess the efficiency of the device. Here $\psi_{\uparrow}(E)$ and $\psi_{\downarrow}(E)$ stand for the wave function for spin up and spin down electrons at energy $E$, respectively. The dependence of these magnitudes on position has been omitted for clarity. Figure~\ref{fig4}(a) shows a density plot of this magnitude for the same GNR considered in figure~\ref{fig3} and the same energy marked there, so the corresponding wave functions are the ones given in figure~\ref{fig3}(b). We can observe that the spin polarisation is focused mainly in two different regions, namely the center of the GNR and the edges. In figures~\ref{fig4}(b) and \ref{fig4}(c), cross sections at the center and the edges are displayed. In both cases, the polarisation in the GNR regions beneath the ferromagnetic lens reaches values close to $P=1$ and $P=-1$, indicating the high degree of spin polarization. This is better observed 
in figure~\ref{fig4}(d), where the maximum of the spin polarisation at the center and edges is calculated for all the energies within the one mode regime. The polarisation at the center is $P_{\mathrm{max}}\simeq 1$ for all the energies, while the polarisation at the edge is about $0.9$ for the lower energies and increases until it reaches a value of approximately $1$.

\subsection{Lens Radius}

\begin{figure}[htb]
\centerline{\includegraphics[width=0.7\columnwidth,clip=]{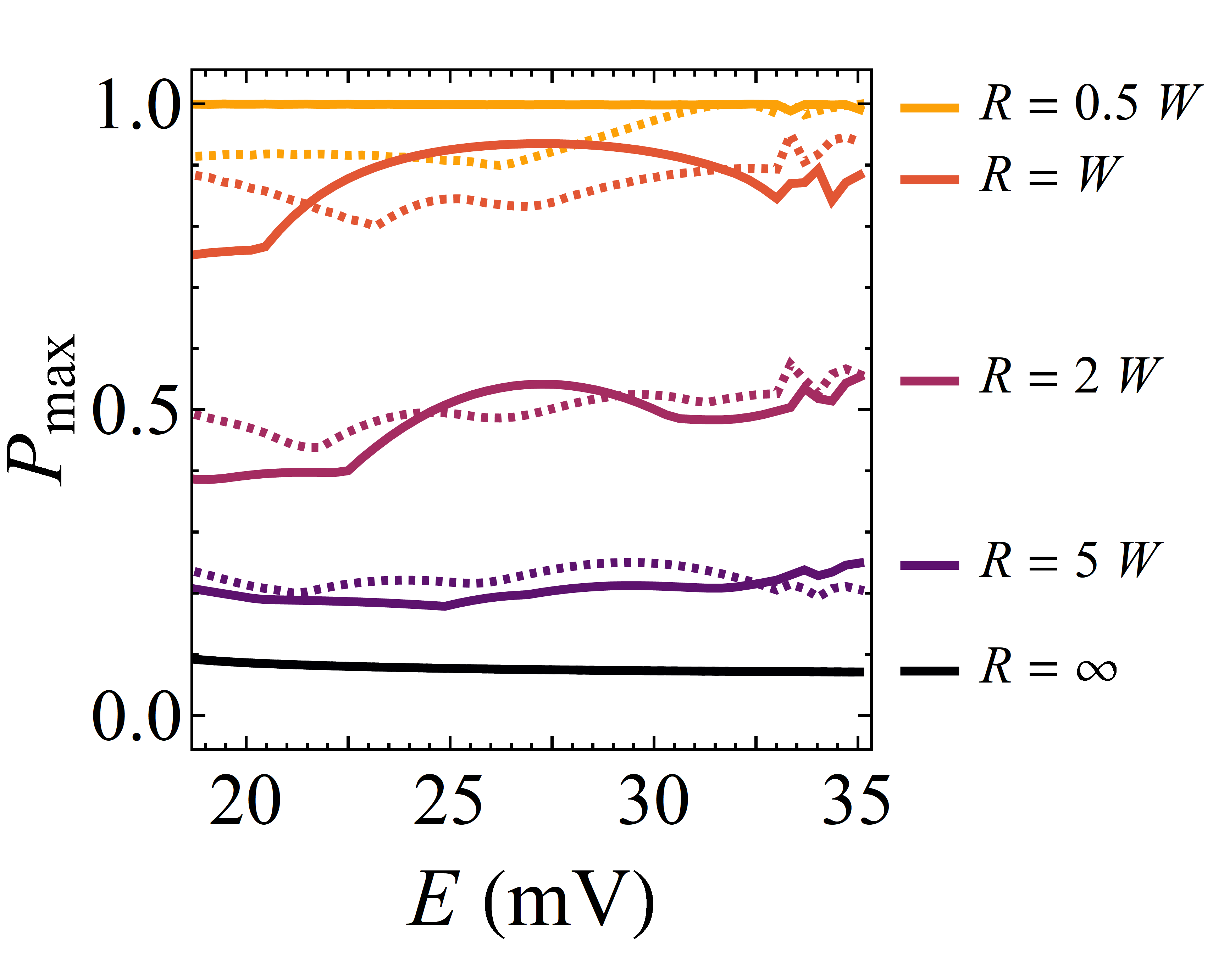}}
\caption{Maximum of the spin polarisation at the center (solid curves) and edges (dotted curves) for all the energies in the one mode regime. A $W=\SI{35}{\nano\meter}$ GNR and different values of the lens radius are considered.}
\label{fig5}
\end{figure}

We also address the role of the lens radius $R$. We fix the GNR width again at $W=\SI{35}{\nano\meter}$ and vary the radius from $R = W/2$, which is the smaller case that can be considered, to $R \to\infty$, which is just a flat lens. In figure~\ref{fig5} we observe that the maximum of the spin polarisation decreases as the radius gets larger, both at the center (solid lines) of the GNR and at the edges (dotted lines). This result demonstrates that the ferromagnetic layer is acting as a lens indeed.

\subsection{Edge Disorder}
\begin{figure}[htb]
\centerline{\includegraphics[width=1\columnwidth,clip=]{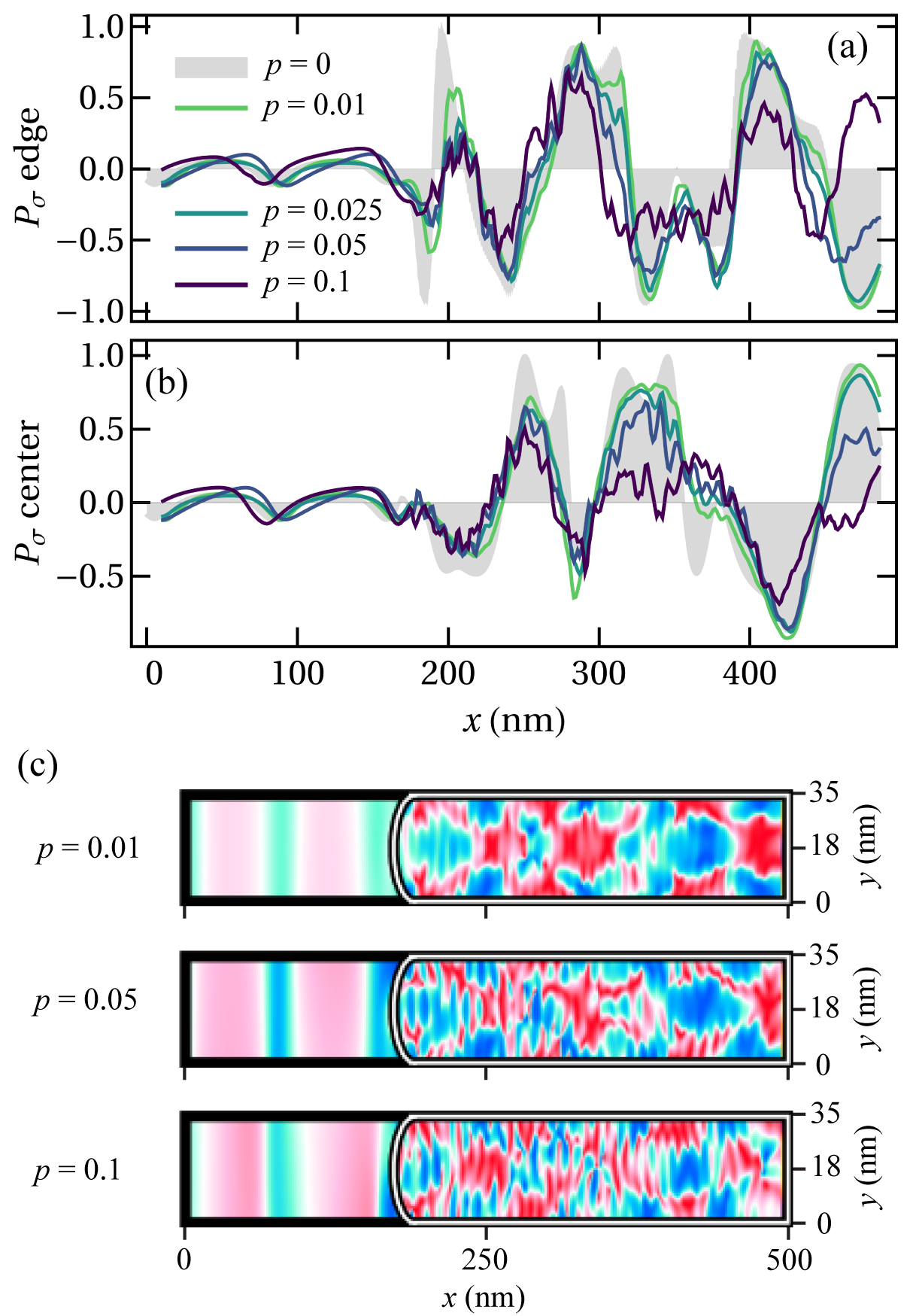}}
\caption{Spin polarisation along (a) the edge, and (b) center, for a GNR of width $W=\SI{35}{\nano\meter}$, radius $R=W/2$, and energy $E=31$~meV (the same case considered in figure~\ref{fig4}). A GNR with perfect edges (gray area) and several values of the probability of removal $p$ are displayed. For each value of $p$, the spin polarisation is averaged over $50$ realizations. (c) Density plot of the spin polarisation for typical realizations of edge disorder. Blue (red) area represents a high magnitude of the spin down (up) density. The probability of removal $p$ is indicated in the plot.}
\label{fig6}
\end{figure}

Atomic size fluctuations, especially at the edges, are unavoidable in real samples. Hence, because the focus spots at the edges are the most promising for device applications, we address the effects of edge disorder on the spin polarisation. 
To do so, we randomly remove pairs of carbon atoms from the edges with some given probability $p$. By removing pairs instead of single atoms, we avoid dangling atoms in the armchair edges, so we do not have to deal with complicated edge reconstructions effects. In figure~\ref{fig6}(a) and (b) we plot the spin polarisation for a GNR with perfect edges (gray area) and several values of the probability of removal $p$. In addition, for each value of $p$, the spin polarisation is averaged over $50$ realizations of edge disorder. We found that when the edge disorder is small ($p\leq 0.025$) the spin polarisation is slightly distorted but not markedly deteriorated. However, as $p$ increases, the spin polarisation is highly degraded. To better visualize these effects, in figure~\ref{fig6}(c) we show the density plot of the spin polarisation for typical realizations of edge disorder. We observe that for $p=0.01$ the density plot is slightly distorted but still resembles the one of the perfect case [see figure~\ref{fig4}(a)], so we conclude our design is robust under moderate edge disorder.

\section{Conclusions}   \label{sec:conclusions}

In summary, we have proposed and studied a novel spin lens which exploits spin-dependent quantum interference effects. The device comprises a GNR and a lens-shaped ferromagnetic layer (e.g. of EuO) grown on top of it. The proximity induced exchange interaction between the magnetic ions and the GNR electrons result in spin-dependent quantum interference effects. A standard two-terminal configuration is used to make electrons flow through the device. In the one-mode regime, which is relevant in narrow GNRs, one of the spins are back reflected while the opposite spins reach the drain with probability close to unity. Nevertheless, the efficiency of the device in this regime is limited by the spin coherence length that needs to be large. This limitation is overcome in wider GNRs, when a large number of electron modes enter the device. The lens-shape edge of the ferromagnet makes electrons focus at the center and edges of the GNR. Most importantly, the spots where the electron density becomes large are different for spin-up and spin-down electrons.
From the standpoint of applications, the edge polarisation is a very attractive and innovative result and also robust under moderate edge disorder, as we demonstrated. In fact, the focalization induced by spherical lenses has been already studied in detail \cite{Cserti2007} but it fails in practical implementations due to the localization in an only one spot inside the lens. In our proposed scenario, a spin-polarised electric current can be easily generated by placing a third non-magnetic contact in the proper place, close to the edge of the ferromagnetic layer.

\acknowledgments
We thank A. V. Malyshev for bringing the attention to this problem and his discussions and comments. We are also thankful to F. Dom\'{\i}nguez-Adame for his support and help in the development of this study. This work was supported by the Spanish MINECO under grant MAT2016-75955.

\bibliography{references}

\end{document}